\documentclass[twocolumn,twoside,slac]{revtex4}
\usepackage{graphicx}
\usepackage{fancyhdr}
\begin{document}

\title{FAMOS: A Dynamically Configurable System for Fast Simulation and Reconstruction for CMS}

%

\author{St. Wynhoff\footnote{current address: CERN, 1211 Geneva 23, Switzerland}}
\affiliation{Princeton University, Princeton, NJ 08544, USA}

\begin{abstract}
Detailed detector simulation and reconstruction of physics objects at
the LHC are very CPU intensive and hence time consuming due to the
high energy and multiplicity of the Monte-Carlo events and the
complexity of the detectors. We present a dynamically configurable
system for fast Monte-Carlo simulation and reconstruction (FAMOS) that
has been developed for CMS to allow fast studies of large samples
of Monte-Carlo events. Each single step of the chain - simulation,
digitization and reconstruction, as well as combinations of chain
links can be replaced by modules that sacrifice precision for
speed. Fast and detailed modules have identical interfaces so that a
change is fully transparent for the user.

Currently, a complete set of the fastest possible implementation,
i.e. going directly from the Monte-Carlo truth to reconstructed
objects, has been implemented. It is about hundred to thousand times
faster than the fully detailed simulation/reconstruction and provides
tracks reconstructed by the inner tracker, clusters in calorimeters
and trigger-Lvl1 objects and tracks reconstructed for the muon system
of the CMS detector.

\end{abstract}

\maketitle

\thispagestyle{fancy}


\section{Introduction}
The detailed simulation and reconstruction of physics events at the
Large Hadron Collider (LHC), $p\bar{p}$ collisions at 14~TeV, where a
large number of high-energy particles are produced and up to 200
collisions overlay, is extremely time consuming even on todays fastest
CPUs. However, to investigate the physics potential of the CMS
detector a large number of collisions must be studied. This is
especially true for scanning the multi-dimensional parameter space of
SUSY models. By replacing the detailed modeling of energy losses and
material effects in simulation as well as the reconstruction
algorithms by simple parameterizations significant improvements
concerning speed can be at the price of precision losses.

\section{The Simulation, Reconstruction and Analysis chain}
The procedure to analyze physics models can be decomposed into four
major steps:
\begin{enumerate}
\item Generation of Monte-Carlo events.\\
  The Monte-Carlo events are created using generators like Pythia,
  Herwig or ISAJET. These generators produce a list of particles -
  stable and decaying - and their four-vectors. In CMS the information
  is stored in form of HBOOK-Ntuples. The time required to generate an
  event is normally less than 100 milliseconds\footnote{All times are
  given as measured on a 1~GHz Pentium III.}.
\item Simulation of material effects.\\ 
  This is the most time consuming step required. Currently, the
  simulation is still done with a Geant3~\cite{g3} based FORTRAN program: {\it
  CMSIM}~\cite{cmsim}. The time varies significantly depending on the number of
  particles and their energies in the Monte-Carlo events. The average
  is between 100 and 200 seconds per event. The information that is
  stored as output of this step are called {\it SimHits}. They contain
  the information about the energy stored in different detector
  elements at different times. A new, object-oriented simulation
  program {\it OSCAR}~\cite{oscar}, based on Geant4~\cite{g4}, is currently being
  validated. 
\item Simulation of readout  electronics (digitization).\\
  The detector converts the energy deposited by the particles into
  electronic signals that are converted to digital information by ADCs
  and TDCs. At the high luminosity at the LHC the detector ``sees''
  the overlap of up to 200 minimum bias events with a single signal
  event. Since the simulation of material effects requires large
  amount of CPU time, the minimum bias events are randomly selected
  from a large pool of simulated events and combined with the
  simulated signal events. So even being technically part of the
  simulation, the combination of minimum bias events with a signal
  event and the simulation of the detector response to the energy
  deposition are performed by the reconstruction software: {\it
  ORCA}~\cite{orca}. The time for this depends on the simulated luminosity and the
  event type. It is 1 to 10 seconds per collision. The output created
  in this step is called {\it DIGIs}. 
\item Reconstruction of physics/analysis objects.\\ 
  The reconstruction is performed by ORCA in several sub-steps. First
  the DIGIs are combined to reconstructed hits, {\tt RecHits}, which
  for example combine several strips of the silicon tracking
  detectors. Similarly, level-1 trigger objects are built. Then
  RecHits are used to find tracks in the inner tracker and the muon
  chambers and clusters in the calorimeters. The reconstruction can
  produce more complicated objects like jets or information about the
  missing energy and finally physics objects like electrons, photons,
  muons etc. The time spent on this can vary largely but is typically
  10 to 100 seconds per collision.
\end{enumerate}

The total CPU time required before the analysis of a collision can be
started is 3-5 minutes. 

\section{Structure of FAMOS}
To give physicists the possibility to study large event sample fast,
the FAMOS\footnote{FAst MOnte-carlo Simulation, see~\cite{famos}} project has been
developed in CMS. The main design concept is to provide fast modules,
able to replace each single step in the chain, as well as several steps
in one go and this complying to the same interfaces as the full simulation and
reconstruction. 

\begin{figure}[hb]
\includegraphics[width=70mm]{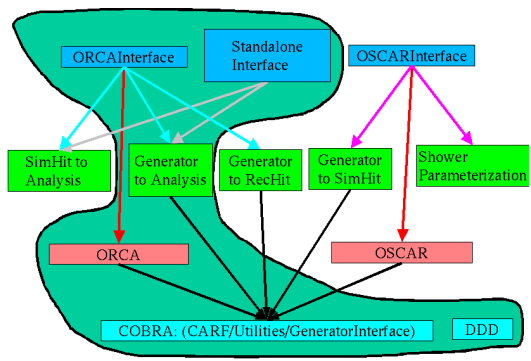}
\caption{FAMOS subsystems and their relation to other CMS
  projects. The arrows indicate the dependencies.}
\label{fig:structure}
\end{figure}

To achieve this FAMOS is divided into several subsystems that focus on
the different steps. Figure~\ref{fig:structure} shows the different
subsystems and their relations to the other CMS object-oriented
software projects. COBRA/CARF~\cite{carf} is the general framework that controls event
handling, DDD~\cite{ddd} provides services for geometry handling, OSCAR is the
simulation software and ORCA the reconstruction software. The
subsystems for a particular task can be accessed via several
interfaces: One providing the same interfaces as for the corresponding
detailed simulation or reconstruction, one for simplified and easy
standalone usage. The implementation of {\it OSCARInterface} and {\it
ORCAInterface} allows to dynamically change from detailed simulation
and reconstruction to the FAMOS implementations.

The first implementation focuses on the replacement of all steps for
simulation and reconstruction and produces physics objects directly
from the Monte-Carlo four-vectors. This corresponds to the hashed area
in Figure~\ref{fig:structure}. The rest of this article describes the
current implementation.

\section{Event handling and simulation modules}
When implementing the direct way from Monte-Carlo events to physics
analysis objects, two main issues arise. First the Monte-Carlo event
is read and second dedicated simulation modules for the different
sub-detectors of CMS are executed.

\subsection{Monte-Carlo event reading}
An important first step in FAMOS is the handling of the original
Monte-Carlo event. The event is stored in the class {\it RawHepEvent}
that is an exact C++ implementation of the {\it HEPEVT} FORTRAN
CommonBlock that is used by most Monte-Carlo generators. An abstract
base class {\it BaseHepEventReader} exists that allows to fill the
{\it RawHepEvent} from a multitude of sources. ASCII files, HBOOK
Ntuples, particle guns - mostly used for single particle tests, from
the databases used as persistency store by OSCAR and ORCA and directly
from the Pythia-6 Monte-Carlo generator. Similar reader modules can be
created for other Monte-Carlo generators. 

The event reader modules are provided by the CMS framework, COBRA, and
used from there. 

\subsection{FAMOS event handling}
\begin{figure*}[tb]
\centering
\includegraphics[width=150mm]{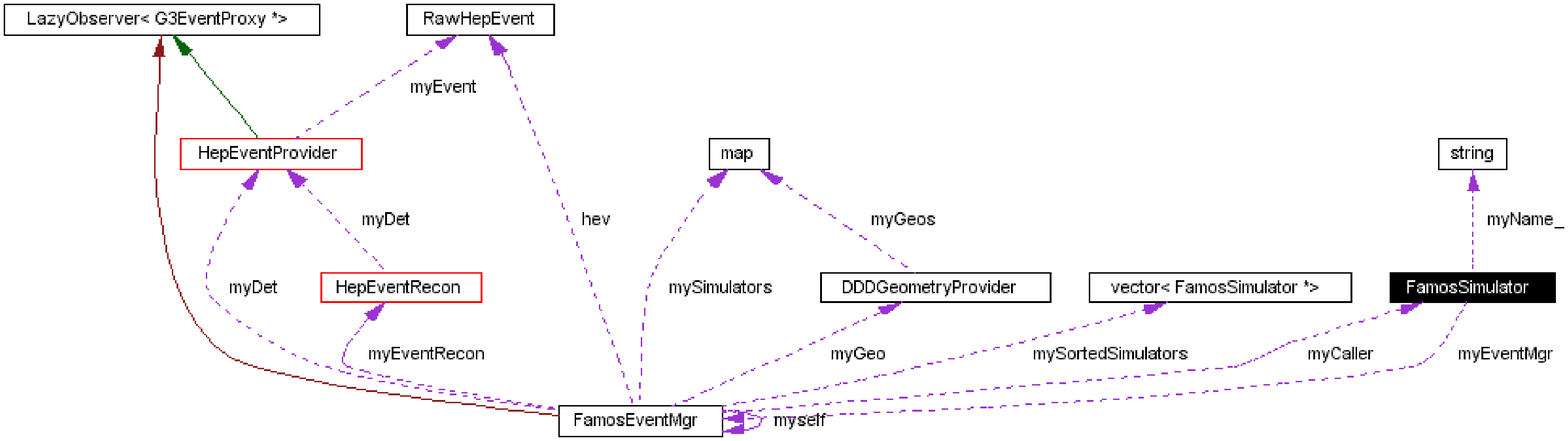}
\caption{Collaboration diagram for the FAMOS event manager class}
\label{fig:eventmgr}
\end{figure*}

In FAMOS an special event manager class, {\it FamosEventMgr}, takes
care of 
\begin{itemize}
\item reading the Monte-Carlo event and
\item calling simulation modules that inherit from {\it FamosSimulator}
\end{itemize}
It interfaces directly to the {\it reconstruction on demand} mechanism
that is a key point of the CMS framework. Figure~\ref{fig:eventmgr} shows a
collaboration diagram between the {\it FamosEventMgr} and the
framework classes. At startup {\it FamosEventMgr} accepts
registrations from the selected {\it FamosSimulator} modules and reads
the geometry. Concrete classes that inherit from {\it FamosSimulator}
implement the fast simulation algorithms.

When starting to loop over events, the {\it FamosEventMgr} first fills
{\it RawHepEvent} from the selected input. {\it RawHepEvent} is also
used to keep track of new particles that can be created by the {\it
  FamosSimulator} classes. This modification of the event is done to
take care of effects like Bremsstrahlung, pair production or multiple
scattering. 

The {\it FamosSimulator} classes are called in a well defined
sequence. It is important to follow the order in which the particles
cross the detector - first the Tracker, then the Calorimeters (ECAL
and HCAL) and only last the Muon system. 

Currently, the simulators are called in the order of their
registration - the user is responsible to do this correctly.

\subsection{Simulation modules and interfaces to the framework}
The classes inheriting from {\it FamosSimulator} are implementing the
concrete fast simulation parameterizations. Currently, implementations
exist for 
\begin{itemize}
\item material effects
\item tracking (Tracker and Muon)
\item electromagnetic clustering
\item muon level-1 trigger
\end{itemize}

The simulation modules - residing in the {\it GeneratorToAnalysis}
subsystem  have to implement a method \\
{\tt bool reconstruct(RawHepEvent \&);}\\
and can add methods to provide the objects they simulate or
reconstruct. All modules provide access to objects that are kept as
simple as possible and a special, framework specific mechanism is then
applied to provide the results in a form compatible with the interfaces
used in normal ORCA jobs: {\tt RecObj}
\begin{figure}[hb]
\includegraphics[width=70mm]{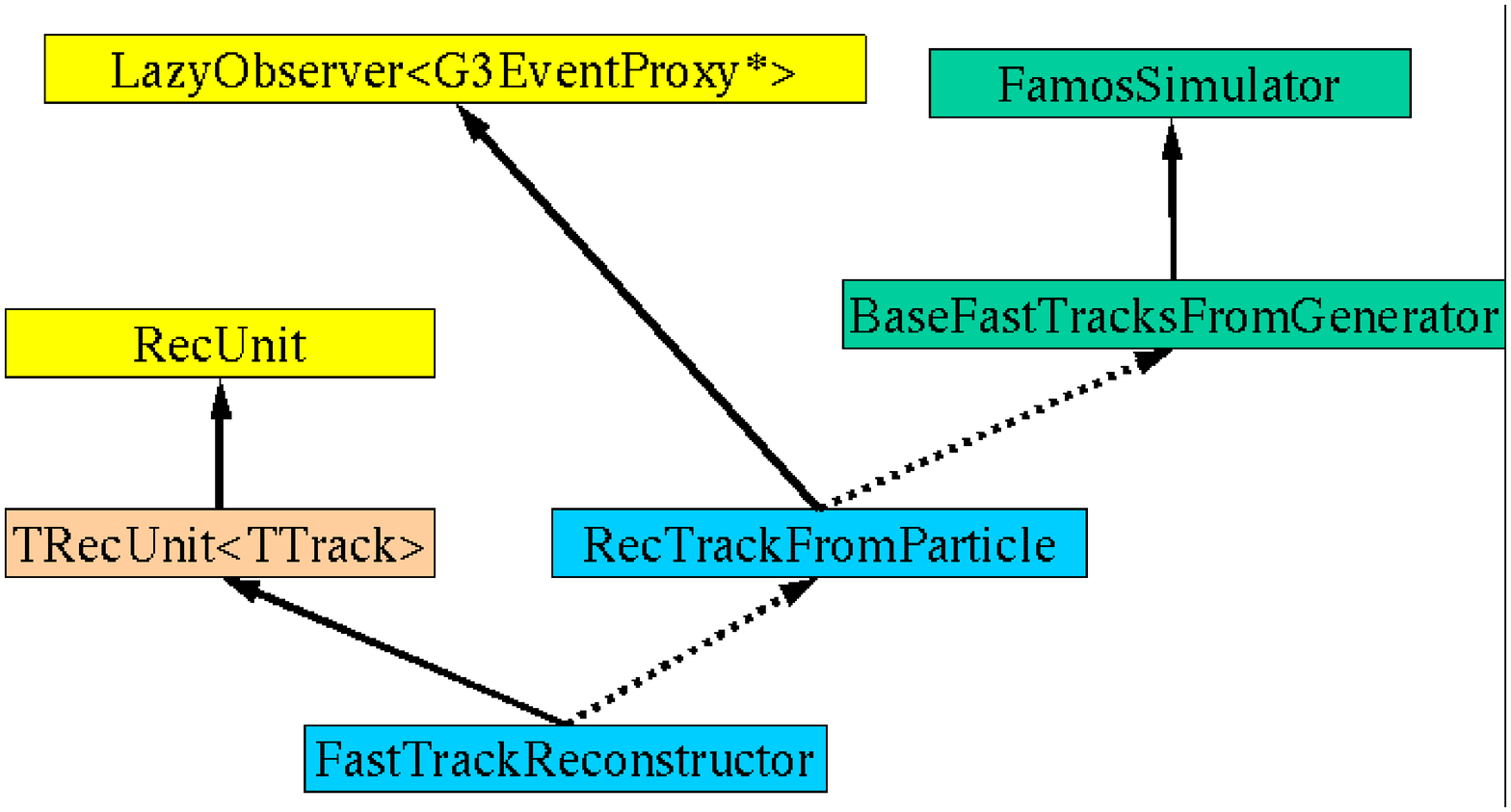}
\caption{The relations between the simulator modules and the ORCA mechanism.}
\label{fig:orcainterface}
\end{figure}

classes. This is done in the {\it ORCAInterface} subsystem. Direct use of the
simulators can be achieved using the {\it StandaloneInterface}
subsystem.

Figure~\ref{fig:orcainterface} shows how classes from COBRA ({\tt
  LazyObserver}, {\tt RecUnit}), ORCA ({\tt TTrack}) and FAMOS
(generic: {\tt FamosSimulator}, {\it GeneratorToAnalysis}: {\tt
  BaseFastTracksFromGenerator}, {\it ORCAInterface}: {\tt
  RecTrackFromParticle}, {\tt FastTrackReconstructor}) work together
  to provide Tracker tracks produced by the fast simulation modules
  with identical interface to the ORCA analysis program as tracks from
  the full reconstruction software.

\section{Example configuration and results}
Since FAMOS provides ORCA compliant interfaces no change is required
to the part of code that uses the objects (e.g. TTracks) when changing
from full to fast reconstruction. However, instantiation of the FAMOS
objects and their registration to the framework must be done. For
Tracker tracks to be provided by the {\it FATSIM} simulation module
this looks like the following:
{\small
\begin{verbatim}
myTrackFinder = new RecTrackFromRawParticle(
                  new FATSIM( &myPtEG, &myAngleEG, 
                              &myEff, 
                              myImpactParameterEG));
myTrackReconstructor = new FastTrackReconstructor( 
                          myTrackFinder, "FATSIM");
[...]
RecCollection<TTrack> MyTracks(ev->recEvent(),
                               "FATSIM");
\end{verbatim}
}
The first lines instantiates an ORCA track-finding object with the {\tt
  FATSIM} simulation module as argument. This simulation module uses
internal classes that provide the parameterizations of transverse
momentum, angular distributions and the efficiencies. Then the ORCA
track reconstructor is created an a string - ``FATSIM'' - is used to
identify it. Creating the collections of {\tt TTrack} specifying the
same string will issue the reconstruction {\it on demand}, i.e. when
iterated over the collection.

This can be compared to the registration of the regular
track finding algorithm:
{\small
\begin{verbatim}
myTrackFinder = new TrackReconstructor(
                 new CombinatorialTrackFinder,
                 "FkFTracks");
RecCollection<TTrack> MyTracks(ev->recEvent(),
                               "FkFTracks");
\end{verbatim}
}
Again a track-finder is instantiated and then a collection. For the
collection the ONLY difference is the string to specify which
reconstruction algorithm is to be used when the collection is
accessed. 

\begin{figure}[h]
\includegraphics[width=70mm]{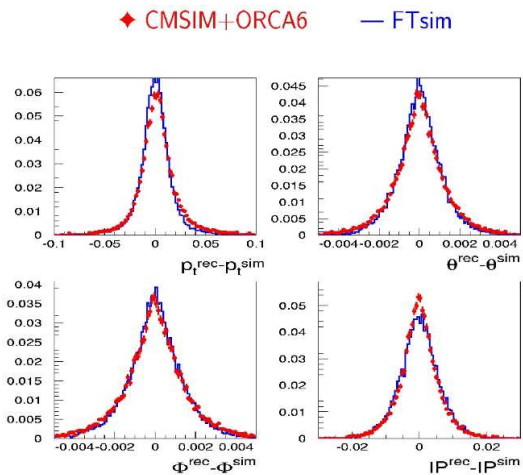}
\caption{Comparison between tracks obtained from full simulation
  (CMSIM) and reconstruction (ORCA6) to tracks obtained from FAMOS for the same
  Monte-Carlo events.}
\label{fig:comparison}
\end{figure}
Similarly, when linking the executable different sets of libraries
must be selected when using full or fast reconstruction. The main
use-case is to run the fast simulation and reconstruction directly
with a Monte-Carlo generator as input. However, it is also possible to
use the FAMOS in parallel to a regular full reconstruction from a
prepared database. This is especially useful for direct comparison
between full and fast simulation and reconstruction. The original
Monte-Carlo event that was simulated in a time-consuming process is
stored in the database, read from there by the {\tt FamosEventMgr} and
then handed to the fast simulators. That allows for example to analyze
two {\tt RecCollections} of {\tt TTrack}s in the same program since
the different collections are mapped via the string identifier to
different track-finding algorithms: detailed reconstruction with {\tt
  RecHit}s as input and FAMOS with the Monte-Carlo event as input.

\section{Timing}
The main aim of the current first implementation was to achieve the
highest possible speed with reasonable agreement. The measured results
for high-multiplicity events without minimum bias pileup is shown in
Table~\ref{tab:timing}. 

\begin{table}[htb]
\begin{center}
\begin{tabular}{|l|r|r|}
\hline  & \textbf{Fraction} & \textbf{time/event [ms]} \\
\hline
Framework       &      3\% &  2 \\
Pythia          &     40\% & 25 \\
FTSim           &      9\% &  6 \\
FastCalorimetry &     35\% & 22 \\
Muon Lvl-1      &      5\% &  3 \\
Muon reconstr.  &      9\% &  6 \\
\hline
CMSIM+ORCA      & 312500\% & 200000 \\
\hline
\multicolumn{2}{|l|}{Material effects prototype} & 2000\\
\hline
\end{tabular}
\label{tab:timing}
\end{center}
\caption{Measured time distribution for FAMOS compared with full
  simulation and reconstruction.}
\end{table}
The FAMOS time is dominated by the Monte-Carlo generator (Pythia) and
the Calorimeter simulation module. All other modules - in particular
the FAMOS framework ({\tt FamosEventMgr}) contribute only
insignificantly. The performance achieved is more than 3000 times
faster than the detailed simulation and reconstruction. 

It is possible to simulate the CMS detector and in particular the
material effects more precisely. A first prototype has been developed
that takes pair production and Bremsstrahlung into account. Without
optimization for performance this reduces the speed of FAMOS to be
only factor 100 faster than CMSIM+ORCA. The modularity - this is just
one additional simulation module to register to {\tt FamosEventMgr} -
give the user full flexibility to adapt the simulation to the
precision his concrete physics analysis requires.

\section{Summary}

FAMOS is a high-performing, flexible and dynamically configurable
mini-framework for fast simulation and reconstruction. It is fully
integrated in the general CMS framework but the components can be used
independently. It is possible to mix full and fast simulation with
minimal changes to the user code. The first modules implemented
provide high-level reconstructed objects (Tracks, Muon, Muon Lvl-1
trigger, Calorimeter clusters) at reasonable agreement in precision
about 3000 times faster than full simulation and reconstruction.

In future fast modules for individual steps in the analysis chain will
be developed and geometry reading will be using the same XML
description as used by the full simulation to ensure consistency.

\begin{acknowledgments}
Special thanks to Marco Battaglia, Filip Moortgat and Artur Kalinowsky
for their implementations and the plots showing example results.
\end{acknowledgments}


\end{document}